# SCIENTIFIC REPORTS

**OPEN**

# Universal scaling of the self-field critical current in superconductors: from sub-nanometre to millimetre size

E. F. Talantsev[1], W. P. Crump[1] & J. L. Tallon[1,2]

Universal scaling behaviour in superconductors has significantly elucidated fluctuation and phase transition phenomena in these materials. However, universal behaviour for the most practical property, the critical current, was not contemplated because prevailing models invoke nucleation and migration of flux vortices. Such migration depends critically on pinning, and the detailed microstructure naturally differs from one material to another, even within a single material. Through microstructural engineering there have been ongoing improvements in the *field-dependent* critical current, thus illustrating its non-universal behaviour. But here we demonstrate the universal size scaling of the *self-field* critical current for any superconductor, of any symmetry, geometry or band multiplicity. Key to our analysis is the huge range of sample dimensions, from single-atomic-layer to mm-scale. These have widely variable microstructure with transition temperatures ranging from 1.2 K to the current record, 203 K. In all cases the critical current is governed by a fundamental surface current density limit given by the relevant critical field divided by the penetration depth.

The key practical property of a superconductor is its maximum dissipation-free current density, the so-called critical current density, $J_c$[1,2]. This quantity varies widely from one superconductor to another and is generally presumed to be governed by the pinning of flux vortices[3]. Over the decades a huge effort has gone into successfully increasing $J_c$ through enhanced pinning[1,3,4]. Pinning, in turn, depends on the detailed microstructure which can vary widely even for a single superconductor. Thus $J_c$ cannot in general be a fundamental property but, rather, is regarded as a materials engineering property. But recently we began to question this conventional view for the special case of zero external field[5] where the only field present is the "self-field" arising from the transport current. We showed[6] for a large range of single- or multi-band superconductors, independent of type, symmetry, size or geometry, that, in the absence of weak links and *in zero external field*, the transport critical current density is determined only by the penetration depth. As noted, this is referred to as the "self-field $J_c$" or $J_c(\text{sf})$. This observation therefore provides a simple route to measure the absolute value of the penetration depth and other thermodynamic properties such as the superconducting energy gap and the relative jump in electronic specific heat at the transition temperature, $T_c$.

Here we explore the implications of this experimental observation. Using a simple but compelling scaling plot we compare $J_c(\text{sf})$ data for nanoscale films and wires, including single-atomic-layer superconductors, with data for macroscopic samples to show that in all cases, spanning up to eight orders of magnitude in effective size, the current density across the surface, when dissipation sets in, is always $J_s = B_c/(\mu_0 \lambda)$ for type I superconductors and $J_s = B_{c1}/(\mu_0 \lambda)$ for type II superconductors. Here $\lambda$ is the London penetration depth, $B_c$ is the thermodynamic critical field:

$$B_c(T) = \frac{\phi_0 \kappa(T)}{2\sqrt{2}\pi \lambda^2(T)}, \tag{1}$$

while, for type II $B_{c1}$ is the lower critical field

[1]Robinson Research Institute, Victoria University of Wellington, P.O. Box 33436, Lower Hutt, 5046, New Zealand. [2]MacDiarmid Institute for Advanced Materials and Nanotechnology, P.O. Box 33436, Lower Hutt, 5046, New Zealand. Correspondence and requests for materials should be addressed to E.F.T. (email: Evgeny.Talantsev@vuw.ac.nz)

   

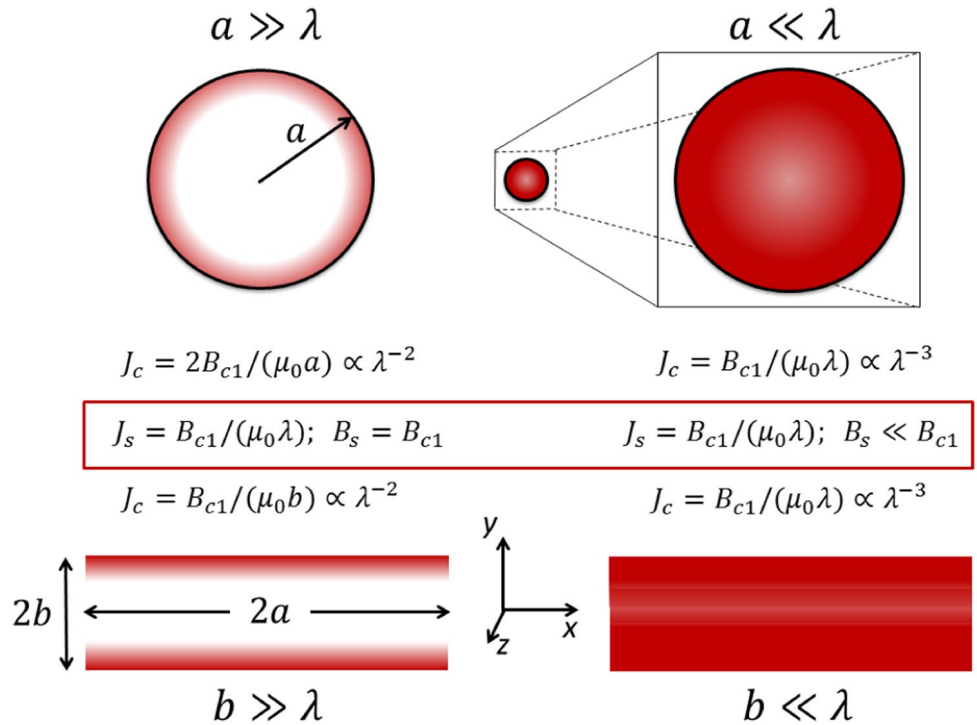

**Figure 1.** The two wire geometries, round wires and thin films, investigated here for self-field $J_c$ showing coordinate system and dimensions. The film cross-section extends from $x = -a$ to $x = +a$ in width, and from $y = -b$ to $y = +b$ in thickness. Depicted are the two extremes when the wire radius, $a$, and film half-thickness, $b$, are either large compared with $\lambda$ (left side) or small compared with $\lambda$ (right side). The displayed formulae summarise our conclusions regarding (i) the global critical current density, $J_c$, averaged over the full cross-section, (ii) the surface current density in the critical state, and (iii) the surface field in the critical state. Asymptotically, for both films or round wires, $J_c(\text{sf}) \propto \lambda^{-3}$ in the small limit and $J_c(\text{sf}) \propto \lambda^{-2}$ in the large limit.

$$B_{c1}(T) = \frac{\phi_0}{4\pi\lambda^2(T)}(\ln \kappa + 0.5), \quad (2)$$

where $\phi_0$ is the flux quantum and $\kappa = \lambda/\xi$ is the Ginzburg-Landau parameter with $\xi$ being the coherence length.

What is compelling about our scaling plot is that calculated surface and edge fields vary extremely widely across all samples investigated (by more than four orders of magnitude) while the surface current density, $J_s$, alone remains universal. This inevitably brings into possible question the conventional picture of dissipation onset due to vortex entry, as we will show.

Figure 1 summarises the two geometries considered - round wires and thin films. Current flows along the $z$-axis and field lines circulate the conductor in the $x$-$y$ plane. For films, the surface current density, $J_s$, and surface field, $B_s$, discussed below are those values on the broad flat surfaces ($y = \pm b$), not the values at the edges ($x = \pm a$). We show schematically the two extremes when the wire radius, $a$, and film thickness, $b$, are either large compared with $\lambda$ (left side) or small compared with $\lambda$ (right side). The displayed formulae summarise our conclusions below regarding (i) the global critical current density, $J_c$, averaged over the full cross-section, (ii) the surface current density, $J_s$, in the critical state, which in all cases adopts the fixed value of $B_c/(\mu_0\lambda)$ or $B_{c1}/(\mu_0\lambda)$, depending on superconductor type; and (iii) the surface field, $B_s$, in the critical state, which only in the large limit is independent of size and equal to the critical field. (We use the term "critical state" to mean at the onset of dissipation when the critical current has been reached. The critical current is customarily defined in terms of a 1 $\mu$V/cm electric field criterion but because we are relying largely on historical literature data not every group has used that criterion though it has now become the norm.). In the small limit $B_s$ falls off as $\tanh(b/\lambda)$. Asymptotically, for both films or round wires, $J_c(\text{sf}) \propto \lambda^{-3}$ in the small limit and $J_c(\text{sf}) \propto \lambda^{-2}$ in the large limit, as will be shown below.

Elsewhere[6], and here, we analyse in total 74 different data sets of $J_c(\text{sf}, T)$ versus $T$ which we fit using extended BCS equations (see Eqs 4 and 5 below), obtaining free-fit parameter values for $\lambda(0)$, $T_c$, $\Delta(0)$ and $\Delta C/C$ where $\Delta(0)$ is the ground-state superconducting energy gap, $C$ is the electronic specific heat and $\Delta C$ is its jump at $T_c$. The purpose of this, in the context of the present work, is solely to enable projection of the $J_c(\text{sf}, T)$ data to $T = 0$ so that we are able to compare ground-state values of critical current density for many different samples across a wide range of size scales. The fact that the parameter values returned by the fits concur with independent measurements gives confidence in the extrapolated values of $J_c(\text{sf}, 0)$ and, further, the returned values of $\lambda(0)$ form a critical test of our model. Samples include metals, alloys, cuprates, pnictides, oxides, nitrides, heavy fermions and borocarbides. In all cases the deduced values of $\lambda(0)$ are in excellent agreement with independently reported values. Most of these previously studied films or wires were of size comparable with $\lambda$ but here we extend the size range considerably to both extremes: $b \ll \lambda$ and $b \gg \lambda$. It is this crucial extension that clarifies to us the precise nature of the phenomenology and leads to the unexpected scaling behavior we report here.

Figure 2 shows reported $J_c(\text{sf}, T)$ values and $\lambda(T)$ values along with their fits deduced using Eqs 3, 4 and 5 below for (a) a single-atomic-layer FeSe film, (b) an almost atomically-thin exfoliated $TaS_2$ film, and, by way of comparison, (c) a relatively large 160 $\mu$m round wire of the Chevrel phase system $PbMo_6S_8$; all of which are s-wave superconductors and, finally, (d) one of our own cuprate $(Y, Dy)Ba_2Cu_3O_7$ films[5]. The cuprate reveals the characteristic d-wave linear temperature dependence of $\lambda(T)$ at low $T$ which is the hallmark of a nodal superconducting gap. In contrast panels (a), (b) and (c) are more consistent with the overall $T$-dependence for s-wave superconductors though it would be preferable to have fuller data sets. Crucially, these examples extend our data range down to single-atomic-layer thicknesses.

The $J_c$ data for these films is processed as described elsewhere[6] and the method is justified rigorously later in this paper. For a type II superconductor when $b \ll \lambda$ then $J_c(\text{sf}) = B_{c1}/(\mu_0 \lambda)$, so

$$J_c(\text{sf}, T) = \frac{\phi_0}{4\pi\mu_0\lambda^3(T)}(\ln\kappa + 0.5), \quad b \ll \lambda \tag{3}$$

Our only input is the experimental value of $\kappa$. For an s-wave superconductor the penetration depth can be expressed in terms of the $T$-dependence of $\Delta$ using[7]

$$\left(\frac{\lambda(0)}{\lambda(T)}\right)^2 = 1 - \frac{1}{2k_BT}\int_0^\infty \cosh^{-2}\left(\frac{\sqrt{\varepsilon^2 + \Delta^2(T)}}{2k_BT}\right)d\varepsilon \tag{4}$$

where $k_B$ is Boltzmann's constant. An analytical expression for the superconducting gap $\Delta(T)$ that allows for strong coupling is given by Gross[8]:

$$\Delta(T) = \Delta(0)\tanh\left[\frac{\pi k_B T_c}{\Delta(0)}\sqrt{\eta\left(\frac{\Delta C}{C}\right)\left(\frac{T_c}{T} - 1\right)}\right] \tag{5}$$

where $\Delta C/C$ is the relative jump in electronic specific heat at $T_c$. For s-wave symmetry $\eta = 2/3$[8]. By combining these equations we fit the experimental $J_c(\text{sf}, T)$ values using $\lambda(0)$, $\Delta(0)$, $\Delta C/C$ and $T_c$ as free fitting parameters. We use non-linear curve fitting in the MATLAB package. For d-wave symmetry the related equations are presented elsewhere[6].

For single-atomic-layer FeSe (Fig. 2(a)) we use the $J_c(\text{sf}, T)$ data set of Zhang et al.[9] along with $\kappa = 72.3$[10]. For $TaS_2$ (Fig. 2(b)) we use the $J_c(\text{sf}, T)$ data of Navarro-Moratalla[11] with $\kappa = 11 \pm 2$. $TaS_2$, like $TaSe_2$, is a multi-band superconductor and this is evidenced by the additional step in $J_c(\text{sf}, T)$ below 0.7 K. For this we use the multi-band fitting analysis developed elsewhere[6]. For $PbMo_6S_8$ (Fig. 2(c)) we use the $J_c(\text{sf}, T)$ data of Decroux et al.[12] with $\kappa = 125$[13]. For this sample with $a \gg \lambda$ Eq. 3 has an additional factor $(2\lambda/a)\tanh(a/2\lambda)$ as applicable to round wires (see below). The $J_c(\text{sf}, T)$ data for YBCO is our own[5] and we use $\kappa = 95$[7]. The green data points on the y-axis are independently measured values of $\lambda(0)$ as follows. For FeSe, from $B_{c1} = 7.5$ mT[10], we infer $\lambda(0) = 324$ nm; for $TaS_2$ reported values vary quite widely - by averaging these we obtain $\lambda(0) = 384 \pm 159$ nm and $\kappa = 11 \pm 2$[11–17]; for $PbMo_6S_8$ $\lambda(0) = 275$ nm[13] and $\kappa = 125$[13]; while for YBCO we use $\lambda(0) = 125$ nm[18], as previously[5,6].

The fit curves are shown in Fig. 2 for $J_c$ (blue curve, right-hand axis) and $\lambda$ (red curve, left-hand axis). They are seen to be excellent (though more complete data sets would be preferable) and as noted the inferred values of $\lambda(0)$ are in very good agreement with independently-measured bulk values (green data points). In particular for FeSe we find $\Delta(0) = 3.0 \pm 0.2$ meV and $\lambda(0) = 336 \pm 2$ nm c.f. 324 nm measured. For exfoliated $TaS_2$ with $2b = 4.2$ nm we find $\Delta_1 = 0.34 \pm 0.05$ meV and $\Delta_2 = 0.14 \pm 0.07$ meV, a two-band gap ratio very similar to that found in $TaSe_2$[19]. The inferred partial band contributions to $J_c$ and $\lambda$ are shown by the black curves. The deduced composite value $\lambda(0) = 394$ nm compares well with the measured $384 \pm 159$ nm. For $PbMo_6S_8$ we find $\Delta(0) = 2.2 \pm 0.2$ meV and $\lambda(0) = 284 \pm 5$ nm c.f. 2.4 meV[20] and 275 nm[13], measured independently. And for YBCO, as discussed elsewhere[6], we find $\lambda(0) = 123$ nm c.f. 125 nm measured independently[18].

The consistency of these fits and the realistic fitting parameters nicely endorse our analysis. It is however important to note that in the scaling analysis to follow we only use these fits to determine the ground-state values ($T \to 0$) of $J_c(\text{sf})$ and $\lambda$. As such the scaling analysis is not at all critically dependent on the detail of these fits.

Having established a reliable method to extrapolate to ground-state values we repeated this analysis for 74 different data sets for widely different superconductors and we turn now to our scaling analysis of this ground-state data. Figure S2 in the Supplementary Information (SI) shows values of $J_c(\text{sf}, T = 0)$ plotted versus sample half-thickness, $b$, or radius, $a$ for many different superconductors, as annotated. There is considerable scatter in $J_c(\text{sf})$ over several orders of magnitude, though for large $b$ a roughly $1/b$ falloff in $J_c(\text{sf}, 0, b)$ is already evident. We wish to scale the dimension of the conductor and the natural length scale to do this is the London penetration depth, $\lambda$. Figure S3 shows the resultant $J_c(\text{sf}, 0, b)$ plotted versus $b/\lambda$. It is probably fair to say that the scatter now seems even greater, with e.g. doped $BaFe_2As_2$ being a significant outlier. At the extreme left is our single-unit-cell FeSe film with $b = 0.275$ nm while the adjacent $TaS_2$ sample is the exfoliated film of thickness 2.1 nm which were discussed in detail in Fig. 2.

The next step is to scale $J_c(\text{sf}, 0)$. The units of current density are the same as those of magnetic field strength divided by a distance, so the natural units of current density for a type I superconductor are $B_c/(\mu_0\lambda)$, and $B_{c1}/(\mu_0\lambda)$ for a type II superconductor. Of course this choice is guided by the outcome but it is the obvious choice in view of our previous work[5]. Thus the normalised critical current density, $J_c^n$, is

$$J_c^n = J_c \times \frac{1}{B_c/(\mu_0 \lambda)}, \tag{6}$$

for type I and

$$J_c^n = J_c \times \frac{1}{B_{c1}/(\mu_0 \lambda)},\qquad(7)$$

for type II.

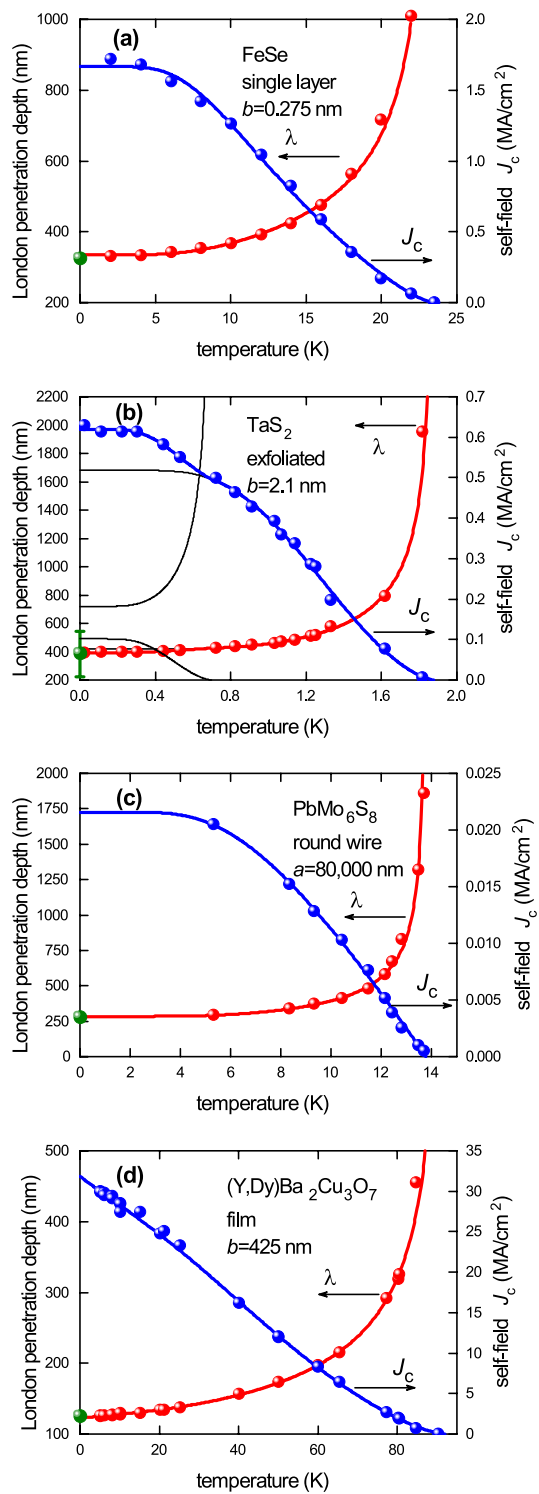

**Figure 2.** (**a**) Experimental self-field $J_c(T)$ data for single-atomic-layer FeSe (right axis, blue) together with values of $\lambda$ (left axis, red) derived as described in the text. The solid curves are the BCS-like fits using Eqs 3, 4 and 5. The single green data point at $T = 0$ is the independently reported ground-state value of $\lambda(0) = 324\,\text{nm}$[10]. (**b**) Presents the same data for ultra-thin exfoliated $TaS_2$ with $2b = 4.2\,\text{nm}$; (**c**) shows the same for a comparatively large round wire of $PbMo_6S_8$ with $a = 80\,\mu\text{m}$; and (**d**) shows the same for $(Y, Dy)Ba_2Cu_3O_7$ from our own work.

We stress here that to calculate $B_c/(\mu_0\lambda)$ and $B_{c1}/(\mu_0\lambda)$ we do not use the values of $\lambda$ that we deduce from our fit procedures. That would be *petitio principii* - circular reasoning. We use independently-measured values of $\lambda(0)$ and $\kappa$ (as listed in Table 1) to calculate $B_c$ and $B_{c1}$ using Eqs 1 and 2. Note also that wherever possible we do not use magnetically-measured values of $B_{c1}$ as these are fraught with experimental uncertainty as we discussed previously[5].

Figure 3 captures our key result. We plot (black symbols) the normalised ground-state conductor critical current density, $J_c^n$, versus the normalised conductor dimension, $b/\lambda$, for the same comprehensive set of superconductors as were discussed in Figs S2 and S3 of the SI. Astonishingly, the broad scatter of Figs S2 and S3 is now almost completely eliminated and in particular the above-noted outlier $BaFe_2As_2$ is entirely consistent now with all other superconductors. The solid curve is the function $(\lambda/b)\tanh(b/\lambda)$ which we previously introduced on purely intuitive grounds[5] but which we rigorously justify later in this paper.

Data for $YBa_2Cu_3O_y$ (YBCO) is shown by the gold symbols. YBCO is anisotropic and this anisotropy needs to be accommodated in the scaling plot for samples with $b > \lambda$, indeed the scaling plot enables the anisotropy to be accurately determined but for the time being we set this to one side. Anisotropy is discussed in more detail later. For all samples numerical values and data sources are listed in Table 1.

Two regions in our scaling plot are evident: For $b < \lambda$ the data points remain essentially constant across the entire small-$b$ range with $J_c^n = 1$; while, for $b > \lambda$ the data falls off as $\lambda/b$ and, as noted, this fall-off was already evident in the raw data (see SI, Figs S2 and S3). We may draw the following key conclusions:

i. This $\lambda/b$ fall-off informs us immediately that the transport current at $J_c$ is not distributed across the full cross-section but is confined to the surface in London-Meissner currents with "dead-space" between. This observation contrasts sharply with the flux-entry model of, for example, Brandt and Indenbom[22] where the current density at $J_c$ is distributed over the entire cross-section due to the ingress of flux vortices. This $\lambda/b$ fall-off is clear evidence against this apparently well-established model. Indeed, the fact that type II superconductors follow the same generic behaviour as type I superconductors in this scaling plot is rather suggestive that vortices (which must be absent in the case of type I) might possibly not play a role in type II superconductors for the onset of *self-field* dissipation. On the other hand it is entirely consistent with London-Meissner transport currents which are certainly present in the case of type I superconductors. Indeed, for type I it is the surface confinement of these London-Meissner currents that is solely responsible for the $\lambda/b$ falloff when $b \gg \lambda$.

ii. For $b < \lambda$ there is also crucial information to be found in the fact that $J_c^n(\text{sf})$ remains constant and equal to unity. The most direct inference, and our firm prediction, is that the current distribution over the entire width, $2a$, must be uniform at the onset of dissipation. The reason for this is that our samples range widely in aspect ratio, $a/b$, from 1 to $3 \times 10^6$. Further, the *effective* aspect ratio given by the ratio of the Pearl length[23], $\lambda_P = \lambda^2/b$ to the half-width, $a$, ranges from $3.5 \times 10^{-3}$ to $1.3 \times 10^7$ with the ratio for one third of our data sets falling less than unity. For these a uniform distribution is expected. But for the remainder, the prevailing if not universal view is that the current must be non-uniformly distributed with the local $x$-dependent current density, $j(x)$, averaged over the film thickness, given by:

$$j(x) = \frac{I}{2\pi b \sqrt{a^2 - x^2}},\qquad(8)$$

as proposed by Rhoderick and Wilson[24] and frequently reiterated since[22,25]. However, our observed scaling behaviour, over such a huge range of effective aspect ratios, precludes such non-uniformity. $J_c^n$ is independent of $a/b$ or $a/\lambda_P$. Moreover, Fig. 3 also includes round wires (to be discussed in more detail later) where, in self-field, the surface azimuthal current density is certainly uniform around the circumference. This huge range of actual and effective aspect ratios all yielding $J_c^n(\text{sf}) = 1$ when $b < \lambda$ or $J_c^n(\text{sf}) = \lambda/b$ when $b > \lambda$ would appear to have no other explanation except that the current density in the films is uniform across the width. We deduce that under self-field the conductor is in a global critical state with the local surface current density saturating everywhere at a critical value, to be identified below as having a magnitude given by the critical field divided by $\lambda$, irrespective of the magnitude of $b$. This critical state begins at the edges and with increasing conductor current the critical region simply extends deeper and deeper into the conductor until, at dissipation, it extends over the full conductor width.

In summary, over the entire range of $b$, and despite the huge variation in $J_c$, sample size and material type, all $J_c^n$ data points lie on a single scaling curve given by $(\lambda/b)\tanh(b/\lambda)$ for films or $(2\lambda/a)\tanh(a/2\lambda)$ for round wires. The $\lambda/b$ fall-off when $b \gg \lambda$ confirms that the transport current is simply the London-Meissner current and the constancy of the amplitude of these scaling functions shows that all these conductors reach a critical state at $J_c(\text{sf})$ with a fundamental surface current density limit given by the critical field divided by $\lambda$; this current density is uniform across the conductor width.

This last conclusion, namely the inference of uniform surface current density at critical current, is so central to our scaling hypothesis that we proceeded to measure the field distribution across the width of commercial RBCO coated-conductor tapes under self-field transport conditions. (RBCO refers to RBa$_2$Cu$_3$O$_7$ where is R in general a mixture of rare earth elements and/or Y). Figure 4 shows our preliminary measurements of the local perpendicular magnetic field, $B_\perp(x)$, over the surface of (a) a 10 mm wide Fujikura tape and (b) a 12 mm wide Superpower tape. We used a cryogenic seven Hall sensor array (Arepoc THV-MOD) with sensors positioned 1.5 mm apart and approximately 0.5 mm above the RBCO film. The data is plotted as $B_\perp/I$ where $I$ is the total transport current which is increased in steps up to the respective self-field transport critical current according to an electric field criterion of 1 μV/cm. The red arrows indicate increasing current. Plotted also is the calculated field distribution for a uniform current distribution (solid curve) and for the Rhoderick and Wilson non-uniform current distribution (dot-dash curve)[24]. It is evident from the data that at low current (20–60 A) the field distribution is indeed consistent with the Rhoderick and Wilson non-uniform current distribution. But, crucially, at higher currents the field distribution deviates from this increasingly and has crossed over, at critical current, to a field distribution fully consistent with the uniform current distribution that we deduced from our scaling analysis. This correspondence is supported both in the shape and in the absolute magnitudes of the local field - with no fitting parameters. In panel (a) the disparity at the extreme left between data and models reflects a small asymmetry in alignment or in conductor performance. In panel (b) all measured fields are slightly higher than the calculated curves but an almost perfect match is achieved if the effective RBCO film width within the tape is 11.7 mm rather than the nominal 12 mm. Many subsequent measurements reinforce this picture in great detail. It is significant also to note that in the original experimental study used by Rhoderick and Wilson[24] for a 20 nm thick Pb strip the current used was about 1/500$^{th}$ of the critical current. Consistent with our data shown in Fig. 4 it is not surprising that at such a low current the non-uniform current distribution was observed. We expect that, at 500 times this current density, Pb films or strips will likewise exhibit a uniform current distribution across the conductor width.

| Material | 2a (nm) | 2b (nm) | κ (expt) | λ(0)(expt) (nm) | $J_c(0)$(MA/cm$^2$) | λ(0)(derived) (nm) | $J_s^n$ | $B_s^n$ | (b/λ) |
|---|---|---|---|---|---|---|---|---|---|
| FeSe (single-atom-layer) | 1.5 mm | 0.55 | 72[10] | 324[10] | 1.65[9] | 336 | 0.897 | 0.000762 | 0.00085 |
| TaS$_2$ (exfoliated) | 450 | 4.2 | 11 ± 2 | 338 ± 143 | 0.62[11] | 394 | 0.631 ± 0.45 | 0.00391 ± 0.0004 | 0.0062 ± 0.002 |
|  |  |  | 13.6, 12.1[16] | 410[17] |  |  |  |  |  |
|  |  |  | 9.8[15] | 302, 260[16] |  |  |  |  |  |
|  |  |  | 9.5[14] | 613[14] |  |  |  |  |  |
| PbMo$_6$S$_8$(round-wire) | 160 μm | — | 125[13] | 275[13] | 0.0216[12] | 284 | 0.936 | 0.936 | 291 |
| NbN (film) | 8.9 μm | 8 | 40[7] | 194[32] | 7.9[33] | 193.5 | 1.0076 | 0.0208 | 0.0206 |
|  | 300 | 8 |  | 200[7] | 14.3[33] | 191.5 | 1.035 | 0.0213 | 0.0206 |
|  | 6.0 μm | 22.5 |  |  | 7.47[34] | 198.6 | 0.9348 | 0.0541 | 0.058 |
| Al (Thin film) (nanowires) | 610 | 89 | 0.03[7] | 50 ± 10[35] | 4.44[21] | 49.3 | 1.0368 | 0.738 | 0.89 |
|  | 680 | 98 |  |  | 3.38[21] | 53.8 | 0.831 | 0.625 | 0.98 |
|  | 500 | 34 |  |  | 3.82[21] | 55.8 | 0.859 | 0.281 | 0.34 |
|  | 300 | 20 |  |  | 3.68[21] | 59.3 | 0.628 | 0.124 | 0.20 |
|  | 10 | 5 |  |  | 9.23[36] | 49.3 | 1.040 | 0.0520 | 0.05 |
|  | 8.4 | 5 |  |  | 7.94[36] | 51.9 | 0.8944 | 0.0447 | 0.05 |
| Ba(Fe, Co)$_2$As$_2$ | 500 μm | 500 μm | 90[37] | 284 ± 15[37–39] | 0.0078[40] | 259 | 1.201 | 1.201 | 880 |
|  | 6.7 μm | 220 |  |  | 2.19[41] | 316 | 0.738 | 0.273 | 0.387 |
| Nb | 82 μm | 1 μm | 1[42] | 49 ± 6[43–45] | 5.12[46] | 50.9 | 0.928 | 0.928 | 10.2 |
|  | 49 μm | 1 μm |  |  | 5.92[46] | 51.2 | 1.064 | 1.064 | 10.2 |
| YBa$_2$Cu$_3$O$_y$ | 500 μm | 850 | 95[7] | 125[18] | 31.8[5] | 122.6 | 1.040 | 1.037 | 3.4 |
|  | 500 μm | 1.4 μm |  |  | 26[47, 48] | 124.3 | 1.011 | 1.011 | 5.6 |
|  | 50 μm | 50 |  |  | 30[49] | 131 | 0.897 | 0.177 | 0.2 |
|  | 5.0 μm | 150 |  |  | 28.9[50] | 134 | 0.902 | 0.484 | 0.6 |
| (STI tape) | 500 μm | 4.5 μm |  | 137 (22 K) | 10.6[48] | 138 | 0.984 | 0.984 | 16.4 |
| (single xtal) | 2 mm | 30 μm |  | 125[18] | 2.04[30] | 123 | 1.032 | 1.032 | 120 |
| (single xtal) | 2 mm | 30 μm |  | 125[18] | 1.50[30] | 143 | 0.761 | 0.761 | 120 |
| Nb$_3$Sn | 101 μm | — | 22[7] | 65[7] | 0.56[51] | 57.7 | 1.27 | 1.27 | 777 |
| (cyl) | 94 μm | — |  |  | 0.46[51] | 66 | 0.972 | 0.972 | 723 |
| (film) | 150 μm | 36 μm |  |  | 0.761[51] | 65.2 | 0.992 | 0.992 | 277 |
| (commercial) | 1.3 cm | 8.5 μm |  |  | 2.59[52] | 65.4 | 0.9904 | 0.9904 | 65.4 |
|  | 1.3 cm | 5.76 μm |  |  | 4.58[52] | 59.7 | 1.186 | 1.186 | 44.3 |
| MgB$_2$ | 320 | 10 | 26[53] | 90[54, 55] | 121[56] | 84.9 | 1.171 | 0.065 | 0.0555 |
|  | 5 μm | 10 |  |  | 78.2[57] | 86.7 | 1.119 | 0.0621 | 0.0555 |
|  | 350 | 100 |  |  | 84.1[58] | 94.2 | 0.889 | 0.448 | 0.556 |
|  | 1.2 mm | 200 μm |  |  | 0.085[59] | 82.2 | 1.376 | 1.376 | 1216 |
| In | 360 | 100 |  |  | 41.1[60] | 34.1 | 1.163 | 0.987 | 1.25 |
| (round-wire) | 520 μm | — | 0.11[7] | 40[7] | 0.0105[61] | 39.1 | 1.072 | 1.072 | 3250 |
|  | 270 μm | — |  |  | 0.0211[61] | 38.5 | 1.119 | 1.119 | 1688 |
|  | 170 μm | — |  |  | 0.036[61] | 37.6 | 1.202 | 1.202 | 1063 |
|  | 170 μm | — |  |  | 0.0347[61] | 38.1 | 1.158 | 1.158 | 1063 |
| MoN (round) | 160 | — | 54[62] | 440 ± 40[64] | 0.5923[63] | 463 | 0.8676 | 0.156 | 0.1818 |
| MoGe | 10 μm | 200 | 94 ± 14[65, 66] | 400[67] | 1.75[65] | 332–339 | 1.731 | 0.424 | 0.25 |
|  | 20 μm | 200 |  |  | 1.31[65] | 366–373 | 1.296 | 0.317 | 0.25 |
|  | 25 μm | 200 |  |  | 1.13[65] | 384–392 | 1.118 | 0.274 | 0.25 |
|  | 30 μm | 200 |  |  | 1.04[65] | 395–403 | 1.029 | 0.252 | 0.25 |
|  | 40 μm | 200 |  |  | 1.02[65] | 397–405 | 1.009 | 0.247 | 0.25 |
| H$_2$S (155 GPa) | 80 μm | 650 (est) | 88[68] | 163[68] | 10.3[68] | 189 | 0.682 | 0.657 | 1.99 |

**Table 1.** Materials, input parameters, source references and derived values of λ(0), the normalized surface current density, $J_s^n$, and normalized surface field, $B_s^n$, at self-field critical current for all data points in Fig. 5. The column for (b/λ) refers to films while for cylindrical samples refers to (a/λ).

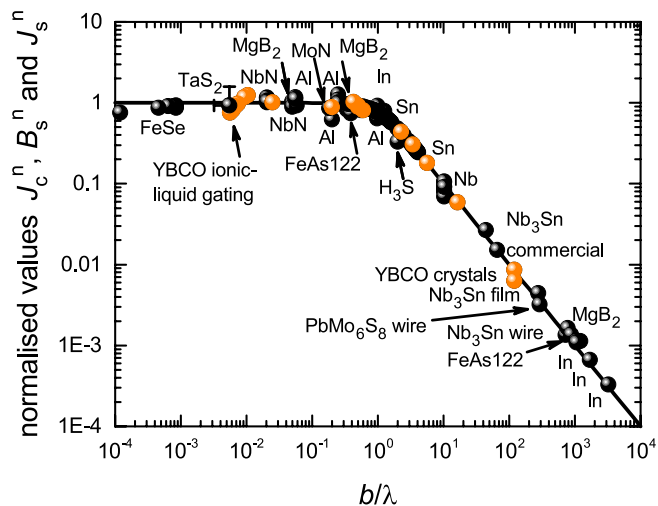

**Figure 3.** Normalised values of the ground-state critical current density, $J_c^n$ (black) for various type I and type II superconductors in films or round wires plotted versus normalised conductor dimension $b/\lambda$ for films and $a/2\lambda$ for round wires. Values are calculated using Eqs 6 and 7 for type I and type II, respectively. Data for $YBa_2Cu_3O_y$ (YBCO) is shown by the gold symbols and because it is anisotropic this data is plotted versus $b/\lambda_c$ where $\lambda_c$ is the c-axis penetration depth taken here to have the value $\lambda_c(0) = 1000$ nm. Anisotropy is discussed in more detail later. $H_3S$ is $H_2S$ compressed to 155 GPa having the current record $T_c = 203$ K[68]. $J_c$ data for $H_3S$ is processed in ref. 69. The error bars for $TaS_2$ are due to excessively disparate reported measurements of $\lambda(0)$ for this system, as described in the text. For this the errors in $J_s^n$ are correlated with those in $(b/\lambda)$ as shown by the sign of the partial error bars.

**We turn now to explicitly quantify our scaling hypothesis using a London analysis.** Consider first our rectangular film illustrated in Fig. 1 of width, $2a$, extending in the x-direction and thickness $2b$ extending in the y-direction with current flowing in the z-direction. As noted, in the critical state at $J_c(sf)$ the local current density $j(x)$ is uniform across the width (except very close to the edges) and the London solution to the self-field current distribution is then x-independent with $J(y) = J_s \cosh(y/\lambda)/\cosh(b/\lambda)$ where $J_s$ is the surface current density. Integrating from $-b$ to $+b$ to obtain the total current then dividing by $2b$ to get the global current density we can express the critical surface current density, $J_s$, in terms of the overall $J_c$ as follows:

$$J_s = J_c \times \frac{(b/\lambda)}{\tanh(b/\lambda)}, \tag{9}$$

From London's second equation the surface field (at $y = \pm b$) at this critical current is

$$B_s = \mu_0 b J_c. \tag{10}$$

As before, we normalise $J_s$ using the natural units of current density of $B_c/(\mu_0\lambda)$ for type I or $B_{c1}/(\mu_0\lambda)$ for type II superconductors. Similarly, we normalise $B_s$ in Eq. 10 by $B_c$ or $B_{c1}$, respectively. Thus e.g. for a type II superconductor the *normalised* surface current density at the onset of dissipation is

$$J_s^n = J_c \times \frac{(b/\lambda)}{\tanh(b/\lambda)} \frac{4\pi\mu_0 \lambda^3}{\phi_0(\ln\kappa + 0.5)}, \tag{11}$$

while the *normalised* surface field at dissipation onset is:

$$B_s^n = \mu_0 b J_c \times \frac{4\pi\lambda^2}{\phi_0(\ln\kappa + 0.5)}. \tag{12}$$

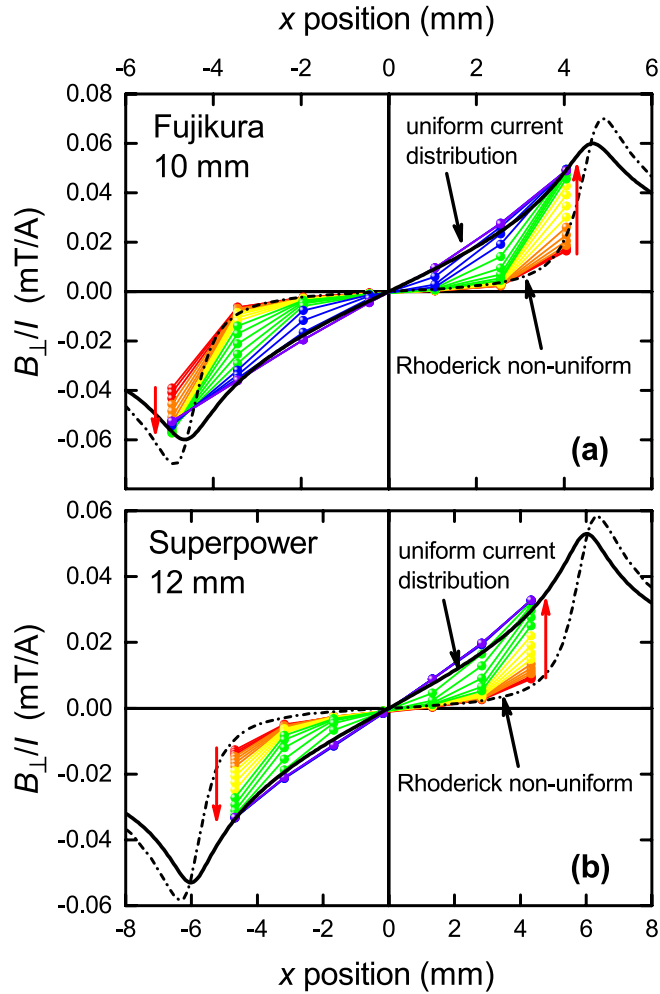

**Figure 4.** The perpendicular field distribution, $B_\perp(x)$, across commercial RBCO tapes measured using a 7 Hall-sensor array for (**a**) a 10 mm wide Fujikura conductor with current increasing from 20 A to 460 A in steps of 20 A; and (**b**) a 12 mm wide Superpower conductor with current increasing from 25 A to 425 A in steps of 25 A. In both cases increasing current is indicated by the red arrows. The sensors are arranged in a line 1.5 mm apart. The black dot-dash curve is the calculated field distribution (0.5 mm above the tape surface at the Hall sensors) for the Rhoderick and Wilson non-uniform current distribution[24] while the solid black curve is the calculated field distribution for a uniform current distribution. As current increases there is a clear crossover from the non-uniform to the uniform current distribution when critical current is attained. The data is plotted as $B_\perp/I$ so as to more clearly expose the detailed evolution at lower currents.

The equivalent expressions for type I will be obvious. Our approach then is to fit the experimental $J_c$(sf, $T$) data over a broad temperature range using the modified BCS equations as above. For example, the fit for single-layer FeSe is shown by the blue curve in Fig. 2(a). By extrapolating to $T = 0$ we obtain the ground-state value $J_c$(sf, $T = 0) = 1.67$ MA/cm$^2$. We use the geometrical factors reported with the data ($b = 0.275$ nm) then use independently reported $\lambda(0)$ and $\kappa$ values to calculate the normalisation factors. The results for single-layer FeSe are $J_s^n = 0.895$ and $B_s^n = 7.5 \times 10^{-4}$. The former shows that the surface current density is close to the mooted critical value $B_{c1}/(\mu_0 \lambda)$, and indeed the reported value of $\lambda(0) = 324$ nm need only be 3.6% larger to obtain $J_s^n = 1$, exactly. This is because of the cube dependence on $\lambda(0)$ and is probably less than the error bars on the value of $B_{c1}$ used to obtain $\lambda$. In contrast $B_s^n = 7.5 \times 10^{-4}$ shows that the surface field is three orders of magnitude less than the bulk lower critical field and, as we show in the SI, at elevated temperatures closer to $T_c$ the surface field is four orders of magnitude less than $B_{c1}$.

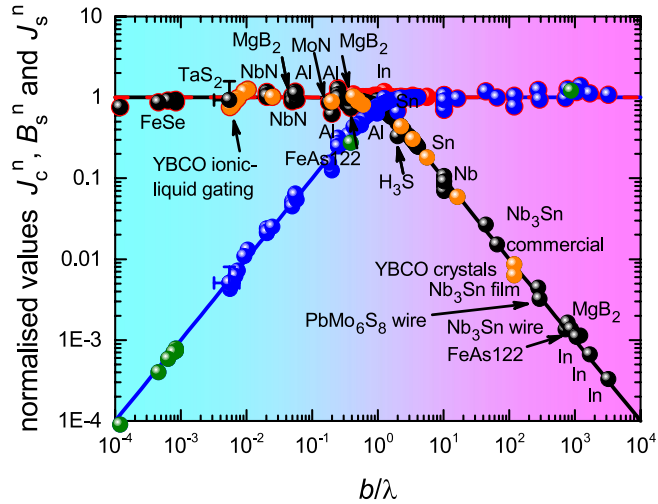

**Figure 5.** Normalized values of the ground-state critical current density, $J_c^n$ (black), surface current density, $J_s^n$ (red), and surface field, $B_s^n$ (blue), for type I and type II superconductors in films or round wires plotted versus normalized conductor dimension $b/\lambda$ for films and $a/2\lambda$ for round wires. Values are calculated using Eqs 11 and 12, 13 and 15. For anisotropic superconductors $\gamma B_s^n$ is plotted instead of $B_s^n$, where $\gamma = \lambda_c/\lambda_{ab}$ –see text. When $b \gg \lambda$ the blue $B_s^n$ data points more or less exactly superpose the red $J_s^n$ data points which are made slightly larger to be seen. For $b \ll \lambda$ the black $J_c^n$ data points sit exactly under the red $J_s^n$ data points. The figure shows that for all size scales the critical surface current density, $J_s$, always equals the critical field divided by $\mu_0\lambda$. $B_s^n$ for the three iron pnictides is shown in green to identify them across the entire range.

Before turning to other materials we first extend the above equations to cylindrical symmetry which will also be used in our analysis of some literature examples. The case of cylindrical symmetry was first discussed by London and London[6, 26]. The local current-density distribution at radial position $r$ in a cylindrical wire of radius $a$ carrying a total current $I$ along its axis in the $z$-direction is $J(r) = [I/(2\pi a)] \times [I_0(r/\lambda)/I_1(a/\lambda)]$, where $I_0(x)$ and $I_1(x)$ are the zeroth- and first-order modified Bessel functions of the first kind. This leads to the following normalised expressions equivalent to Eqs 11 and 12:

$$J_s^n = J_c \times \frac{a}{2\lambda} \frac{I_0(a/\lambda)}{I_1(a/\lambda)} \times \frac{4\pi\mu_0\lambda^3}{\phi_0(\ln\kappa + 0.5)} \quad (13)$$

$$\approx J_c \times \frac{(a/2\lambda)}{\tanh(a/2\lambda)} \times \frac{4\pi\mu_0\lambda^3}{\phi_0(\ln\kappa + 0.5)}, \quad (14)$$

and

$$B_s^n = \frac{1}{2}\mu_0 a J_c \times \frac{4\pi\lambda^2}{\phi_0(\ln\kappa + 0.5)}, \quad (15)$$

together with the equivalent expressions for type I superconductors. Note that in these equations there is no requirement for local current density uniformity around the circumference as was the case for films. Previously we found the approximation to the ratio of Bessel functions in Eq. 14 was excellent across the entire range of $a$ values and is exact at both the large and small asymptotes. In our calculations we prefer to use the exact Bessel function expression but the correspondence between Eqs 11 and 14 is notable and shows that $J_c(\text{sf})$ versus $a/(2\lambda)$ for round wires has the same scaling as $J_c(\text{sf})$ versus $b/\lambda$ for rectangular films. This also suggests a means of addressing the case where the film width, $2a$, is finite. In particular we replace the second factor in Eq. 11 by:

$$(\lambda/b)\tanh(b/\lambda) \to (\lambda/a)\tanh(a/\lambda) + (\lambda/b)\tanh(b/\lambda). \quad (16)$$

This is our only approximation; it is used only for finite-width films and, because of the observed uniform current distribution at $J_c$, we believe it to be a good approximation to within a factor close to unity. When $a \gg \lambda$ it reduces to our exact expressions and the approximation is certainly good enough when we consider that both the current and the dimension range over some seven orders of magnitude. When $a = b$ and $b \ll \lambda$ it reduces to the cylindrical case.

We see that Eqs 11 and 14 immediately capture the scaling behaviour shown in Fig. 3, namely $J_c^n = J_s^n \times (\lambda/b)\tanh(b/\lambda)$ for films and $J_c^n = J_s^n \times (2\lambda/b)\tanh(b/2\lambda)$ for round wires with in either case $J_s^n = 1$. This shows explicitly that at the onset of self-field dissipation the surface current density is always equal to the critical field divided by λ. Figure 5 reproduces the $J_c^n$ scaling plot (black symbols) and to it we now add the normalised surface field and surface current density. This summmarises all our results. Details are as follows.

Given the ground-state $J_c(sf, 0)$ data and using Eqs 12 or 15 we plot the normalized critical-state surface field, $B_s^n$ (blue data points) as it varies with normalised conductor dimension. The solid blue curve is the scaling function $B_s^n = \tanh(b/\lambda)$ which arises naturally from Eqs 11 and 12 if $J_s^n = 1$. Finally, using Eqs 11 or 13 we plot the ground-state normalized surface current density, $J_s^n$, (red data points) calculated from the $J_c(sf, 0)$ data. These red data points are made slightly larger than the blue data points so that they can be seen behind them when $b > \lambda$. Across the entire size range the scaled parameter $J_s^n$ remains close to unity (red dashed line) showing that in every case, small or large, the surface current density at $J_c(sf)$ is indeed given by the relevant critical field divided by $\mu_0\lambda$. $B_s^n$ also shows a common scaling behavior. The three $B_s^n$ or $J_s^n$ data points for the ferro-pnictides, single-atomic-layer FeSe and Ba(Fe, Co)$_2$As$_2$, are shown in green firstly to identify them where there is a cluster of data points, but also to highlight the two extremes of behavior at the nanoscale and macroscale in a single class of superconductor, as well as an intermediate case.

Figure 5 reveals a common universal scaled behavior namely that, whatever the size, dissipation sets in when $J_s$ reaches $B_c/\mu_0\lambda$ for type I and $B_{c1}/\mu_0\lambda$ for type II superconductors, and this condition occurs uniformly across the entire conductor surface. This is fully understandable for type I because this corresponds to the London depairing current density, $J_d$. But for type II $J_d$ is higher by a factor of $\sqrt{2}\kappa/(\ln\kappa + 0.5)$ ($\approx 27$ for YBCO) and the result is new and puzzling. Importantly, only for large dimensions with respect to λ does this threshold coincide with $B_s$ reaching the relevant critical field. For small dimensions the surface field falls far short of the bulk critical field, indeed even below Earth's magnetic field. The dissipation threshold is thus not field-limited but, according to Ampere's law, *field-gradient-limited* corresponding to $(\partial B_s/\partial y)_{y=\pm b} = J_s = B_c/(\mu_0\lambda)$ or $(\partial B_s/\partial y)_{y=\pm b} = J_s = B_{c1}/(\mu_0\lambda)$.

The case of anisotropic superconductors is discussed in detail in the SI. The scaling curve for the normalised $J_c$ is given in this case by:

$$J_c^n = (\lambda_c/b)\tanh(b/\lambda_c). \quad (17)$$

where $\lambda_c$ is the *c*-axis penetration depth and the crystallographic *c*-axis of the film is parallel to the *y*-axis of the coordinate system shown in Fig. 1. As a consequence identical scaling behavior should be found as was found for the isotropic superconductors provided that $(b/\lambda_c)$ is plotted on the *x*-axis. This was done for YBCO in Figs 3 and 5 so as to bring all superconductors together in a single plot. Note that the effects of anisotropy vanish when $b \ll \lambda_c$. Our fit procedure does allow for the anisotropy parameter $\gamma = \lambda_c/\lambda_{ab}$ to be extracted as a free-fitting parameter (where, for a layered superconductor, $\lambda_{ab}$ is the in-plane penetration depth, usually abbreviated to just λ). However, this scaling behavior immediately suggests an alternative approach to directly determine the magnitude of $\gamma$, something that historically has been difficult to pin down. If we plot $J_c^n$ versus $(b/\lambda_{ab})$ (instead of $(b/\lambda_c)$) using a log-log plot, then we obtain the scaling curve as for the isotropic superconductor but rigidly displaced to the right by the factor $\gamma$. We illustrate this in Fig. 6 for the case of YBCO. One simply needs good quality self-field $J_c$ data for a wide range of film thicknesses (in which of course the *c*-axis is normal to the plane of the film). The data in the literature is rather limited in this regard but nonetheless sufficient to illustrate the method. From Fig. 6 we obtain $\gamma = 7$, very consistent with for example the value determined by torque magnetometry[29]. Note also that $B_s^n$ (blue data points) is displaced upwards by the factor $\gamma$ as shown in the figure. Hopefully this work will prompt the accumulation of a wider data range in order to determine the anisotropy with more accuracy, and potentially as a function of doping where we expect it will change rapidly.

As shown in the SI, for anisotropic superconductors the surface field only scales like the isotropic superconductors when plotting $\gamma^{-1}B_s^n$ versus $b/\lambda_c$. This is what was done in Fig. 5 for YBCO. This then brings together films and round wires, for both isotropic and anisotropic superconductors into a single scaling plot for $J_c^n$ and $B_s^n$.

What are the physical origins of this apparently universal scaled behavior in type II superconductors? Importantly we should be guided by the fact that this behavior is completely analogous to that exhibited by type I superconductors. It is as though the current density $B_{c1}/\mu_0\lambda$ acts as some kind of depairing limit in type II just as $B_c/\mu_0\lambda$ *is* the depairing limit in type I. For the case of thin films we may consider just three possibilities: (i) conventional vortex entry from the edges[22], (ii) vortex entry from the flat surfaces as we proposed earlier[5], or (iii) a non-vortex, and hence new, mechanism.

**Vortex entry at the edges.** The edge field, $B_y$, can be calculated using Eq. (4) of Brojeny and Clem[25] who consider the case of a uniform current distribution over the film cross section. These are the conditions that certainly pertain when $b \ll \lambda$. Brojeny and Clem express the field in terms of the current density. At $J = J_c$ we may replace $J_c$ by $J_s$ using Eq. 9 (recall that $b \ll \lambda$) and then replace $J_s$ by $B_{c1}/(\mu_0\lambda)$ to obtain the edge field:

$$B_y(x = \pm a) = B_{c1}(1/\pi)\tanh(b/\lambda)\left[\ln(\frac{2a}{b}) + 1\right]. \quad (18)$$

Even though the assumption of uniform current density restricts this expression nominally to $b \ll \lambda$ we believe it to be general, for all $b$, including where for $b \gg \lambda$ the current is confined just to the London screening layer at the surface. To see this we consider the approximation of replacing the wire of rectangular cross-section, $2a \times 2b$, by a planar assembly of round wires, each of radius $b$, placed side by side to a width of $2a$ and carrying the same total current. The field at the edge can be calculated exactly for all $b$. We find (see SI)

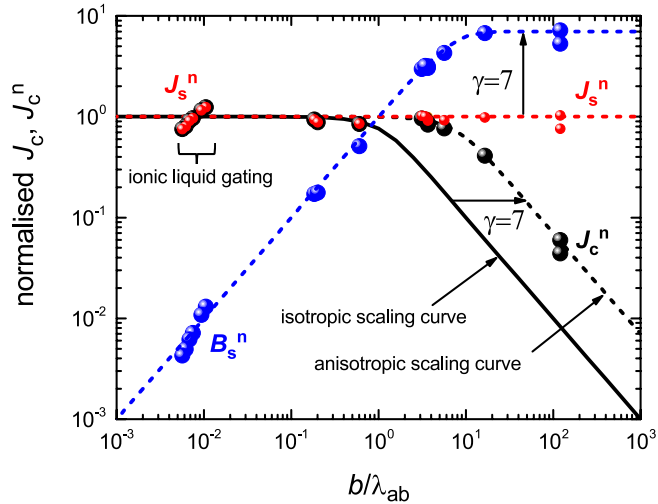

**Figure 6.** Illustrating the method to measure the anisotropy parameter $\gamma = \lambda_c/\lambda_{ab}$ for YBa$_2$Cu$_3$O$_7$ (YBCO) films and single crystals. Normalized values (black symbols) of the critical current density, $J_c^n$, are plotted versus normalized conductor dimension $b/\lambda_{ab}$. The solid black curve is the scaling curve for isotropic superconductors and is identical to the black curve in Fig. 5. The black dashed curve is the solid curve displaced along the $x$-axis by the factor $\gamma$. By extrapolating the data on the sloping ($1/b$) arm back to $J_c^n = 1$ the value of $\gamma$ may be read off as the abscissa. Here we obtain $\gamma = 7$. Parenthesised data at small $b$ is for a single 4.5 nm thick film of YBCO at different doping levels achieved by ionic-liquid gating[27]. Values of $\lambda(0)$ are obtained from the doping-dependent superfluid density[28]. The two data points at the large $b$ are for two 30 $\mu$m thick single crystals[30] and the next highest $b$ value is for a 4.5 $\mu$m thick STI tape measured in our laboratory. The remainder are films reported in Table 1. The blue data points are the scaled surface field, $B_s^n$ and blue dashed curve is the scaling curve given in the SI by Eq. S7, $B_s^n = \gamma \tanh(\gamma^{-1} b/\lambda_{ab})$. The red data points are the scaled surface current density, $J_s^n$ and the red dashed line is the scaling curve $J_s^n = 1$. The red data points are made smaller so as to make the black data points under them visible.

$$B_y(x = \pm a) = B_{c1}(1/\pi)\tanh(b/\lambda)\left[\ln(\frac{2a}{b}) + 1.2704\right], \quad (19)$$

which is almost identical to Eq. 18 which we take therefore to be valid for all $b$. In the SI we discuss the effect of anisotropy. The modification to Eq. 18 is quite straightforward and leads to:

$$B_y(x = \pm a) = \gamma B_{c1}(1/\pi)\tanh(b/\lambda_c)\left[\ln(\frac{2a}{b}) + 1\right] \quad (20)$$

$$\rightarrow B_{c1}(1/\pi)(b/\lambda_{ab})\left[\ln(\frac{2a}{b}) + 1\right] \quad b \ll \lambda_c \quad (21)$$

$$\rightarrow \gamma B_{c1}(1/\pi)\left[\ln(\frac{2a}{b}) + 1\right] \quad b \gg \lambda_c \quad (22)$$

The first difficulty to note with the edge field is that, in the conventional picture of vortex entry from the edges, this dependence of the edge field on the ratio $a/b$ would preclude our observed scaling behavior. In the examples of Fig. 3 the ratio $a/b$ ranges over six orders of magnitude. Even though $a/b$ sits within a logarithm the factor $[\ln(2a/b) + 1]$ still ranges from 1 to 16.6. Within this picture the edge field simply could not scale with $(b/\lambda)$, yet $J_c^n$, $\gamma^{-1}B_s^n$ and $J_s^n$ all do so, without exception.

Secondly, the edge field ranges from 0.0045 $B_{c1}$ to 12.1 $B_{c1}$ and shows no correlation whatsoever with $J_c(\text{sf})$. A key player here is the anisotropy factor $\gamma$ in Eq. 20 which can lead to very large values of the edge field relative to $B_{c1}$ when $b \gg \lambda_c$. When $b \ll \lambda_c$ the parameter $\lambda_c$ cancels out and then the edge field can be extremely small because of the resulting $(b/\lambda_{ab})$ factor.

This is our third point: it can be seen that for $b \ll \lambda$ the edge field is always much smaller than the bulk $B_{c1}$. Consider single-atom layer FeSe where we find $B_y = 0.0045 B_{c1}$ at $J_c(\text{sf})$ or exfoliated TaS$_2$ where $B_y = 0.0133 B_{c1}$.

Fourthly, where the edge field does exceed $B_{c1}$ (when $b > \lambda$) we can then ask how deep the field penetrates before it falls to the value of $B_{c1}$. In the presence of strong pinning once $B_y$ does reach $B_{c1}$, as shown by Brandt and Indenbom[22], the field falls abruptly to zero and there is no current density in the interior of the conductor beyond this point (except for the London current at the surface). Again we can use Brojeny and Clem[25] to calculate the maximal extent of the flux entry at the edges. Rearranging their Eq. (13) (or indeed their Eq. (8) for a finite-thickness film) and generalising to the anisotropic case, the *y*-component of the field at any point on the axis from $x = -a$ to $x = +a$ becomes

$$B_y(x) = \frac{\gamma B_{c1} \tanh(b/\lambda_c)}{2\pi} \ln\left[\left(\frac{x+a}{x-a}\right)^2\right]. \tag{23}$$

The amplitude $\gamma B_{c1}\tanh(b/\lambda_c)$ is just the value of the *x*-component of the normalized surface field discussed above and shown in Fig. 5. For $b \ll \lambda$ this field is tiny. But for $b \gg \lambda$ consider as an example the case of the thick MgB$_2$ film in Table 1 with $2b = 0.2$ mm. Inserting these numbers and using $\lambda(0) = 90$ nm[54,55] we find that $B_y$ falls to the value of $B_{c1}$ in just $0.043 \times 2a$ from the edges and this is essentially negligible. The same applies for each of the few very thick films listed in Table 1.

Fifthly, and perhaps more important than any of the above, we obtain precisely the same scaling behavior (with $a/2\lambda$) for the *round* cross-section examples in Table 1 where there is no singularity at the edges that might drive early entry of vortices. Our observed scaling behaviour is entirely consistent with the London-Meissner model and inconsistent with conventional vortex models. In this regard we also recall the previously reported insensitivity of $J_c(\text{sf})$ to pinning enhancement[5].

**Vortex entry at the flat surfaces.** In our previous work[5] we postulated that vortices might enter the flat surfaces. That work was largely confined to cases where $b \leq \lambda$ and here of course the nucleation, ingress and annihilation of vortices is energetically favoured by the logarithmically-diverging interaction between vortices entering opposing faces. Further, for isotropic superconductors with $b \lambda$ the surface field is found to always just reach the magnitude of $B_{c1}$, whether for highly-aspected films or for round wires. But, for anisotropic superconductors, the surface field can substantially exceed the critical field. So we have precisely the same quantitative problem here as with vortex entry at the edges. For $b < \lambda$ at $J_c(\text{sf})$ we find $B_s = B_{c1}\tanh(b/\lambda_{ab})$, even in the anisotropic case, so here the surface field falls far below the bulk critical field. Now it may be that the effective critical field, $\widetilde{B}_{c1}$, for small samples varies as $\widetilde{B}_{c1} = B_{c1}\tanh(b/\lambda)$ (or for round wires $\widetilde{B}_{c1} = B_{c1}\tanh(b/2\lambda)$) i.e. it is diminished in proportion to its size. This was proposed by Brandt[22] for the effective lower critical field of a very thin film in a longitudinal external field. If this were the case then for all small samples we would have the surface field just reaching the effective critical field for both isotropic and anisotropic superconductors. We remain open to this possibility. But there are difficulties. Firstly, as noted, thick anisotropic samples at $J_c$ can have a surface field which far exceeds the critical field. Secondly, at the other extreme for the single-atomic layer FeSe sample this would require $\widetilde{B}_{c1} \approx 6.4\,\mu\text{T}$. This is as low as one tenth earth's magnetic field. This would mean that the sample on cooling should already be threaded with flux from the earth's ambient field. Thirdly, when $b/\lambda = 0.00085$, as in this example, any conventional description of a vortex fails completely, still less its nucleation at the surface and annihilation with vortices of the opposite sense arriving from the opposing surface. Indeed because here $b \approx 0.06\xi(0)$ there cannot even be a conventional vortex core structure. *x*-oriented vortices in such a confined environment cannot exist. And yet the same $J_c$ scaling behavior found in macroscopic samples is observed in these atomic-scale superconductors with $J_s = B_{c1}/\lambda$. Fourthly, the range of superconductors investigated here must represent an extremely varied spectrum of pinning profiles. Any kind of vortex model would reflect this variable pinning in a distribution of $J_c(\text{sf})$ magnitudes. This does not appear to be observed. Lastly, it is perhaps understandable that when $b < \lambda$ any vortices nucleated on opposite faces of the film will strongly attract and annihilate, with the logarithmically-diverging interaction term immediately overcoming any surface barriers. Thus it is arguable that dissipation could immediately set in when $B_s$ reaches $\widetilde{B}_{c1}$. But this will not be the case in macroscopic samples where surface barriers and pinning profiles will have to be overcome by $B_s$ significantly exceeding $B_{c1}$. This is not observed. And again we note the apparent pinning independence of $J_c(\text{sf})$[5].

**Non-vortex mechanisms.** If indeed vortices are absent then a new physical mechanism for the onset of dissipation is probably required to explain our analysis of type II superconductors under self-field conditions. Our results perhaps indicate this possibility. The fact that type I and II superconductors exhibit essentially the same scaling behavior certainly suggests a common non-vortex model, perhaps based on pair-breaking. What might this new mechanism be? Hirsch[31] has postulated the existence of a Meissner *spin current* coexisting with the conventional Meissner *charge current* which, in our case, is the transport current. These spin currents, in the absence of an external field, flow in opposite directions for each up, or down, spin component. In his picture dissipation would set in when the charge current density reaches $B_{c1}/(\mu_0\lambda)$ just as we have deduced. This corresponds to one of the spin currents being reduced to zero. His analysis is for the case $b \gg \lambda$ and so it remains to be seen whether this result is also sustained when $b < \lambda$. This is the only alternative picture that we are aware of for the onset of dissipation that leads to a universal self-field $J_c$ such as we have inferred. Doubtless there may be others. Finally, we ask how our scaling analysis differs from previous ideas regarding universal onset of dissipation under self-field. Bean and Livingston[70] introduced the concept of a barrier to vortex entry arising from interaction of a vortex near the surface with its image vortex. But the penetration field under such a barrier can substantially exceed $B_{c1}$. And for putative vortex entry at the flat surfaces the surface field, $H_x$, when $b > \lambda$ is universally observed to just reach $H_{c1}$ at $J_c$ with no apparent barrier at all. Moreover, for $b \ll \lambda$ the Bean-Livingston barrier falls to zero due to interaction with the second image vortex at the opposing surface and interaction with any opposite-sign vortices nucleated there. It is difficult to see how our universal $J_s$ value could be recovered from a surface-barrier-type model for such a vast range of sample dimensions and also for round wires.

Our results are perhaps more akin to the idea of a so-called 'geometrical barrier' introduced, for example, by Zeldov and co-workers[71] in which the presumed non-uniform current distribution given by their Eq. (1) is responsible for a vortex potential barrier at the edges. The resulting penetration self-field at the edge is given by $H_p \approx H_{c1} \times \sqrt{2b/a}$ at which the local edge-current-density is $B_{c1}/(\mu_0 \lambda)$. This may seem to correspond to our result, however, we stress that we find this local critical current to occur not just at the edges but uniformly over the entire conductor width, independent of absolute aspect ratio or effective Pearl aspect ratio, and independent of pinning or surface and edge quality. Indeed, because of this current uniformity at the onset of dissipation there is no existing geometrical edge barrier of the form described by Zeldov et al.[71].

Perhaps more significantly, Genenko and co-workers consider the self-field entry of flux vortex rings into cylindrical conductors and find this occurs at a universal surface current density of $J_s = \sqrt{2} B_c/(\mu_0 \lambda)$[72] or alternatively vortex loop nucleation when $J_s = \sqrt{2} \mu_0 \lambda)$[73]. The criterion generally applies for all $a/\lambda$. Though these values are much higher than the critical surface current density that we observe these authors state that for real conductors with surface imperfections on a size-scale $\delta$ where $\xi < \delta < \lambda$ then $B_c$ in the above expressions should be replaced by $(\xi/\delta)B_c$. And if $\delta \lambda$ then $B_c$ should be replaced by $B_{c1}$. This then looks like our result and we do not discount the possibility of vortex loop nucleation as described by these authors however we have to note that this model allows for surface current densities anywhere in the range $B_{c1}/(\sqrt{2} \mu_0 \lambda) \leq J_s \leq \sqrt{2} B_c/(\mu_0 \lambda)$, depending on surface quality. For a high $\kappa$ system this admits a variation in $J_s$ of a factor of 50–60. This is simply not observed and moreover the lower limit is too low by a factor $1/\sqrt{2}$. To sum up, we do not see any *consistent* precedent in the literature for our observations that might conclusively identify the physical mechanism behind our observed universal scaling.

In conclusion, by examining a large number of superconductors of both type I and type II over a huge range of sample dimension we show that in all cases the onset of dissipation at the critical current coincides with the surface current density reaching a value given by the critical field divided by $\lambda$: $B_c/(\mu_0 \lambda)$ for type I and $B_{c1}/(\mu_0 \lambda)$ for type II. This is equivalent to a *surface-field-gradient* threshold and not a surface-field threshold as in the Silsbee criterion. For small sample dimensions with $b \ll \lambda$ the surface field falls far short of the critical field - by more than three orders of magnitude in the case of single-atomic-layer superconductors. We argue that under these circumstances flux vortices are likely to not be present and therefore possibly play no role in the dissipation process. Further, we extract from this scaling analysis a simple method to determine the anisotropy factor in anisotropic superconductors. Our deduced current density criterion is obvious enough in the case of type I superconductors: the surface current density simply reaches the London depairing current density and as a consequence the normal state is nucleated at the surface and the transition for the remainder of the sample simply runs away as the critical interface displaces towards the interior. But for type II superconductors we have a surprising new result that mimics the behavior for type I as though $B_{c1}/(\mu_0 \lambda)$ were some kind of effective pairbreaking current density, but only at the surface. This is closely reminiscent of Hirsch's spin-current picture[31]. Whatever the case this observation renders a universal behavior for all superconductors that can at the very least be used to measure rather precisely the absolute magnitude of the penetration depth and superfluid density–even for multi-band and anisotropic superconductors.

JLT thanks the Marsden Fund of New Zealand for financial support. Authors thank Dr. Jian Wang, Dr. Yi Sun, Dr. Ying Xing (International Center for Quantum Materials, School of Physics, Peking University, People's Republic of China), and Dr. Joshua Island (Kavli Institute of Nanoscience, Delft University of Technology, The Netherlands) for providing raw self-field critical current data for single-atomic-layer FeSe film, and exfoliated $TaS_2$ crystals, respectively. We thank Drs N.M. Strickland and S.C. Wimbush for providing critical current data for YBCO wires and Dr. A.E. Pantoja for assistance in setting up the field-profile measurements.

## Data availability

The data and data sources used in this paper are listed in Table 1. A further comprehensive tabulation is also available in ref. 6.

## Author Contributions

E.F.T. conceived the idea of a surface current density limit, collected all the literature data and proposed to use Eqs 4 and 5 to deduce thermodynamic parameters; WPC and E.F.T. developed the modelling software and E.F.T. carried out the data set fits; J.L.T. and E.F.T. conceived the scaling analysis; E.F.T. and J.L.T. conceived idea of measuring field profiles and E.F.T. performed experiments; J.L.T. drafted the manuscript which was revised and edited by all parties.

## Additional Information

**Supplementary information** accompanies this paper at doi:10.1038/s41598-017-10226-z

**Competing Interests:** The authors declare that they have no competing interests.

**Publisher's note:** Springer Nature remains neutral with regard to jurisdictional claims in published maps and institutional affiliations.